\documentclass[11pt]{iopart2}

\usepackage{graphicx,bm,float,amsmath,amssymb}

\makeatletter

\usepackage[english]{babel}
\usepackage[T1]{fontenc}
\usepackage[utf8]{inputenc}

\begin{document}

\title[Monopole-free CP$^{N-1}$ models in the presence of a quartic 
potential]{Three-dimensional 
monopole-free CP$^{N-1}$ models: \\ 
Behavior in the presence of a quartic potential}

\author{Claudio Bonati$^1$, Andrea Pelissetto$^2$ and Ettore Vicari$^1$} 
\address{$^1$Dipartimento di Fisica, Universit\`a di Pisa,
        and INFN, Sezione di Pisa \\ 
        Largo Pontecorvo 3, I-56127 Pisa, Italy}
\address{$^2$Dipartimento di Fisica, Universit\`a di Roma ``La Sapienza",
        and INFN, Sezione di Roma  \\
        P.le Aldo Moro 2, I-00185 Roma, Italy}

\ead{Claudio.Bonati@unipi.it,Andrea.Pelissetto@roma1.infn.it,\\ Ettore.Vicari@unipi.it}

\date{\today}

\begin{abstract}
We investigate the phase diagram and the nature of the phase
transitions in a three-dimensional model
characterized by a global SU($N$) symmetry, a local U(1) symmetry,
and the absence of monopoles. It represents a natural generalization
of the gauge monopole-free (MF) CP$^{N-1}$ model, in which the 
fixed-length constraint (London limit) is relaxed.  We have performed 
Monte Carlo simulations for $N=2$ and 25, observing a
finite-temperature transition in both cases, related to the
condensation of a local gauge-invariant order parameter.  For $N=2$
results for the MF model are consistent with  a weak
first-order transition. A continuous transition would be possible only
if scaling corrections were anomalously large.
For $N=25$ the results in the general MF model are also consistent with 
a first-order transition, that becomes weaker as the size of the
field-length fluctuations decreases.
\end{abstract}

\maketitle


\section{Introduction}
\label{intro}

In recent years it has been realized that the phase diagram and critical
behavior of classical and quantum models with U(1) gauge symmetry 
crucially depends on topological aspects of the model 
\cite{SBSVF-04,MV-04,BS-13,BMK-13,NCSOS-15,WNMXS-17,Sachdev-19}. 
For instance, in the
quantum case, the behavior depends on the presence or absence of the 
Berry phase, while in the classical case the topology of the gauge
configuration space plays a crucial role. In a series of recent papers 
\cite{PV-19-AH3d,BPV-20-AHq2,BPV-22-AHq,BPV-21-NCQED,PV-20-mfcp}
we investigated the role that topological properties of the U(1) gauge fields
play in the multicomponent Abelian-Higgs (AH) model, in which an $N$-component
complex scalar field is coupled to a U(1) gauge field. We studied models
with compact \cite{PV-19-AH3d,BPV-20-AHq2,BPV-22-AHq} and noncompact gauge 
fields \cite{BPV-21-NCQED} and a model \cite{PV-20-mfcp} in which monopoles 
are suppressed---we used the prescription of De Grand and Toussaint 
\cite{DGT-80}. Results confirmed that topology plays an important role, 
both for small and large values of $N$. 

For $N\ge 10$,
the AH model with noncompact gauge fields (NCAH) has a line of continuous 
transitions that appear to be naturally associated with the stable 
charged 
fixed point that occurs in the AH field theory \cite{HLM-74,FH-96,IZMHS-19}.
Along the transition line, both scalar and gauge degrees of freedom 
play a role.
Such a transition line is absent in the compact lattice AH model with 
charge-one scalar fields \cite{PV-19-AH3d}, 
but is present if the scalar fields have an integer
 charge $q$, satisfying $q\ge 2$
\cite{BPV-20-AHq2,BPV-22-AHq}. The different behavior for $q = 1$ and 
$q\ge 2$ is due to the presence or absence of topological transitions. 
For $q=1$ the potential between charge-one static sources always saturates at
large distances, while for $q\ge 2$, ${\mathbb Z}_q$ gauge excitations give
rise to transition lines
separating phases where the charge-one static sources are confined by a linearly
rising potential from one phase in which they are not.
The presence of this type 
of topological phases appears to be a necessary requirement for the 
observation of transitions controlled by the charged fixed points 
of gauge field theories.

For $N=2$, the AH lattice model has a line
of continuous transitions emerging from the CP$^{N-1}$ critical point,
which belong to the
O(3) universality class for any value of the charge $q$
\cite{PV-19-AH3d,BPV-20-AHq2}.  Gauge fields have only the role 
of hindering some degrees of freedom (those that are not 
gauge invariant) from becoming critical and 
the critical behavior can be explained by an effective 
Landau-Ginzburg-Wilson model in terms of a scalar gauge-invariant 
order parameter.
For any $N\ge 3$ only first-order transitions occur 
\cite{PV-19-AH3d,PV-20-largeN}, 
contradicting the large-$N$ analytic 
predictions \cite{DHMNP-81,YKK-96,MZ-03}. 
Note, however, that, at least for $N=3$, there are some indications 
of continuous transitions in other models \cite{NCSOS-11,NCSOS-13} 
that, on the basis of the 
usual symmetry considerations, are supposed to be in the same universality
class as the AH lattice model.

The behavior of the monopole-free (MF) CP$^{N-1}$ model 
defined in Ref.~\cite{PV-20-mfcp} is much less clear. 
For $N=2$ one observes a transition \cite{PV-20-mfcp}, 
which has some features that characterize
first-order transitions---a broad distribution of the order parameter and
absence of finite-size scaling---and some others that instead are specific of 
continuous transitions. For $N=2$ it also shares some qualitative features that 
have been observed in loop models that are expected to belong to the same
universality class \cite{NCSOS-11,NCSOS-13,WNMXS-17}.
For $N=25$ the model apparently undergoes a continuous
transition. In Ref.~\cite{PV-20-mfcp} it was conjectured that this transition
is associated with the charged fixed point that controls the RG flow in the 
AH field theory for large $N$.
However, the recent results of 
Refs.~\cite{BPV-22-AHq,BPV-21-NCQED} 
exclude this possibility, so that there is at present 
no field-theory interpretation for the transition. 
The presence of two distinct continuous transitions in the NCAH model and 
in the compact MFCP$^{N-1}$ model for $N=25$ 
is quite disturbing as it seems to contradict the 
standard assumption that monopoles are the relevant topological excitations
that characterize the phase diagram of the
model. In particular, {\em  a priori} one would have expected 
continuous transitions in 
the MF model and in the
NCAH model---monopoles are absent here, too---
to belong to the same universality class associated with the 
AH field theory. To add to the confusion on the role of monopoles, we should
mention that in compact models with higher-charge scalar fields, 
monopoles---at least with
the De Grand-Toussaint prescription \cite{DGT-80}--- apparently do not play
any role \cite{BPV-20-AHq2}. 

In this paper, we reconsider the problem, investigating the 
behavior of a generalization of the model of Ref.~\cite{PV-20-mfcp}. 
We consider here unconstrained scalar fields, relaxing the condition 
that their modulus should be equal to 1 (London limit), and adding a quartic
potential. Such a change is expected, at least in the standard 
field-theory approach, to be irrelevant for the critical
behavior.  If the transitions observed in Ref.~\cite{PV-20-mfcp} for 
$N=2$ and $N=25$ are continuous, we expect to observe the same critical
behavior for any generic value of the quartic coupling.
If this occurs, this would strengthen the arguments in favor of a continuous 
transition for these two values of $N$.

The paper is organized as follows. In Sec.~\ref{sec2} we define the model.
In Sec.~\ref{sec3} we define the observables that we compute, and report the 
basic finite-size scaling (FSS) results we use to analyze the numerical data.
In Sec.~\ref{sec4} we present our numerical data and in Sec.~\ref{sec5} we draw
our conclusions.

\section{The model} \label{sec2}

We consider a U(1) gauge model with $N$-component scalar fields 
defined on a cubic lattice. The model is invariant under local
U(1) and global SU($N$) transformations.
The fundamental
fields are complex $N$-component vectors ${\bm w}_{\bm x}$,
associated with the sites of the lattice
and complex phases $\sigma_{{\bm x},\mu}$, 
$|\sigma_{{\bm x},\mu}|=1$, associated with the lattice
links.  The corresponding Hamiltonian is
\begin{equation}
    H = H_{\rm kin} + H_{\rm pot}.
\label{Hamiltonian}
\end{equation}
The first term is 
\begin{equation}
H_{\rm kin} = - N J \sum_{{\bm x}, \mu}
\left( \bar{\bm{w}}_{\bm x}
\cdot \sigma_{{\bm x},\mu}\, {\bm w}_{{\bm x}+\hat\mu}
+ {\rm c.c.}\right),
\label{hcpnla}
\end{equation}
where the sum is over all lattice sites ${\bm x}$ and directions $\mu$
($\hat{\mu}$ are the corresponding unit vectors). The second term is 
\begin{equation}
H_{\rm pot} = - g \sum_{{\bm x}} (|w_{\bm x}|^2 - 1)^2,
\end{equation}
which represents a quartic potential for the field. The 
CP$^{N-1}$ model with unit-length fields ~\cite{RS-81,DHMNP-81,BL-81} is 
recovered in the limit $g\to \infty$. The partition function is 
\begin{equation}
Z = \int [d\sigma_{{\bm x}\mu}] [d{\bm w}_{\bm x}d\bar{{\bm w}}_{\bm x}]\,
   e^{-H/T}.
\label{Zdef}
\end{equation}
In the following we will use $\beta = J/T$ and $\lambda = g/T$ as 
independent variables.
One can easily check that the
Hamiltonian (\ref{Hamiltonian}) is invariant under the global SU($N$)
transformations
\begin{equation}
{\bm w}_{\bm x} \to U {\bm w}_{\bm x} ,\qquad U\in {\rm SU}(N),
\label{unsym}
\end{equation}
and  the local U(1) gauge transformations
\begin{equation}
{\bm w}_{\bm x} \to e^{i\alpha_{\bm x}} {\bm w}_{\bm x} , \qquad
\sigma_{{\bm x},\mu} \to 
e^{i\alpha_{\bm x}} \sigma_{{\bm x},\mu}  e^{-i\alpha_{\bm x+\hat{\mu}}} .
\label{u1gausym}
\end{equation}
The lattice CP$^{N-1}$ model, which is obtained for $\lambda \to \infty$,
has a continuous transition for $N=2$ in the O(3)
universality class, while the transition is of first order for any
$N\ge 3$ \cite{PV-19-CP,PV-20-largeN}.  Note that the transition is
not continuous even for $N=\infty$, in disagreement with analytic
calculation \cite{DHMNP-81,PV-19-CP} performed for this lattice 
model (see Ref.~\cite{PV-20-largeN} for a discussion). As we shall 
discuss, in the general model 
with Hamiltonian (\ref{Hamiltonian}) the behavior is analogous: 
we find that for $N=3,4$ the transition is of first order. 

To explore the role that topological defects play, we consider a model in which
monopoles are absent. Monopoles are defined using the De
Grand-Toussaint prescription \cite{DGT-80}.
In this approach one starts from the
noncompact lattice curl $\Theta_{{\bm x},\mu\nu}$ associated with each
plaquette
\begin{equation}
  \Theta_{{\bm x},\mu\nu} = 
    \theta_{{\bm x},\mu} + \theta_{{\bm x} + \hat{\mu},\nu} - 
    \theta_{{\bm x},\nu} - \theta_{{\bm x} + \hat{\nu},\mu},
\end{equation}
where $\theta_{{\bm x},\mu}$ (with $-\pi < \theta_{{\bm x},\mu} \le \pi$)
is the phase associated with $\sigma_{{\bm x},\mu}$, 
$\sigma_{{\bm x},\mu} = e^{i\theta_{{\bm x},\mu}}$. 
Here $\mu$ and $\nu$ are the directions that
identify the plane in which the plaquette lies. Note that
$\Theta_{{\bm x},\mu\nu}$ is antisymmetric in $\mu$ and $\nu$, so that
we associate two different quantities that differ by
a sign with each plaquette. To define a monopole, let us consider an 
elementary lattice cube. We consider each plaquette $P = ({\bm x},\mu\nu)$
($\mu \not= \nu$) 
belonging to the cube, 
ordering $\mu$ and $\nu$ so that $\hat{\mu}\times\hat{\nu}$ 
points outward with respect to the cube. The number of monopoles 
inside the elementary cube is defined as 
\begin{equation}
N_{\rm mono}(C) = \sum_{P} 
    m\left( {\Theta_{{\bm x},\mu\nu}  \over 2\pi} \right),
\end{equation}
where the sum is over all plaquettes belonging to the cube and 
\begin{equation}
   m(x) = x - \left\lfloor x + 1/2 \right\rfloor.
\end{equation}
To define a monopole-free (MF) version of the  model,
we only consider 
configurations such that $N_{\rm mono}(C) = 0$ for each elementary cube.

\section{The observables} \label{sec3}

In our numerical study we consider cubic lattices of linear size $L$
with periodic boundary conditions.  We simulate the system using 
an overrelaxation algorithm.
It consists in a stochastic mixing of
microcanonical and standard Metropolis updates of the lattice
variables.\footnote{To update each lattice variable, we randomly
  choose either a standard Metropolis update, which ensures
  ergodicity, or a microcanonical move, which is more efficient than
  the Metropolis one but does not change the energy.
  In the Metropolis update, changes are tuned so
  that the acceptance is approximately 1/3.} 
When the MF model is simulated, 
if the proposed move generates a monopole, the move is
rejected.

We compute the energy density and the specific heat, defined as
\begin{eqnarray}
E = {1\over N V} \langle H \rangle,\qquad
C ={1\over N^2 V}
\left( \langle H^2 \rangle 
- \langle H \rangle^2\right),
\label{ecvdef}
\end{eqnarray}
where $V=L^3$.

The model we consider, both in the presence and in the absence of 
monopoles, is expected to 
undergo transitions where the global SU($N$) symmetry is broken. 
The corresponding order parameter is 
\begin{equation}
Q_{{\bm x}}^{ab} = \bar{w}_{\bm x}^a w_{\bm x}^b - {1\over N}
|w_{\bm x}|^2 \delta^{ab},
\label{qdef}
\end{equation}
which is a gauge-invariant hermitian and traceless $N\times N$ matrix
that transforms as
\begin{equation}
Q_{{\bm x}} \to {U}^\dagger Q_{{\bm x}} \,{U}
\label{symmetry-U(N)}
\end{equation}
under the global SU($N$) transformations (\ref{unsym}).  

We consider the two-point correlation function 
\begin{equation}
G({\bm x}-{\bm y}) = \langle {\rm Tr}\, Q_{\bm x}^\dagger  
Q_{\bm y} \rangle,  
\label{gxyp}
\end{equation}
the corresponding susceptibility and correlation length,
\begin{eqnarray}
&&\chi =  \sum_{{\bm x}} G({\bm x}) = 
\widetilde{G}({\bm 0}), 
\label{chisusc}\\
&&\xi^2 \equiv  {1\over 4 \sin^2 (\pi/L)}
{\widetilde{G}({\bm 0}) - \widetilde{G}({\bm p}_m)\over 
\widetilde{G}({\bm p}_m)},
\label{xidefpb}
\end{eqnarray}
where $\widetilde{G}({\bm p})=\sum_{{\bm x}} e^{i{\bm p}\cdot {\bm x}}
G({\bm x})$ and ${\bm p}_m =
(2\pi/L,0,0)$. In the FSS analysis we use renormalization-group invariant 
quantities. We consider
\begin{equation}
R_\xi = \xi/L
\end{equation}
and the Binder parameter
\begin{equation}
U = {\langle \mu_2^2\rangle \over \langle \mu_2 \rangle^2} , \qquad
\mu_2 = {1\over V^2} \sum_{{\bm x},{\bm y}} {\rm Tr}\, Q_{{\bm
    x}}^\dagger Q_{\bm y} .
\label{binderdef}
\end{equation}
To determine the nature of the transition, one can consider the 
size dependence of the 
maximum $C_{\rm max}(L)$ of the specific heat.  At a first-order
transition, it behaves as
\begin{equation}
C_{\rm max}(L) = {1\over 4} \Delta_h^2 V \left[ 1 + O(V^{-1})\right],
\label{Cmax-first}
\end{equation}
where $V=L^d$  is the $d$-dimensional volume ($d=3$) and $\Delta_h$ is the latent
heat.  At a continuous transition, instead, we have
\begin{equation}
C_{\rm max}(L) = a L^{\alpha/\nu} + C_{\rm reg},
\end{equation}
where the constant term $C_{\rm reg}$ is due to the analytic
background. It is the dominant contribution if $\alpha < 0$.  Thus, the 
analysis of the $L$-dependence of $C_{\rm max}(L)$ may allow one 
to distinguish first-order and continuous transitions.
However, experience with
models that undergo weak first-order transitions indicates
that in many cases the analysis of the specific heat is not
conclusive. The behavior (\ref{Cmax-first}) may set in at values of
$L$ that are much larger than those at which simulations can be actually
performed.  In the case of weak first-order transitions, a more useful
quantity is the Binder parameter $U$. At a first-order transition, the
maximum $U_{\rm max}(L)$ of $U$ for each size $L$ behaves
as~\cite{CLB-86,VRSB-93}
\begin{equation}\label{Ufirst}
U_{\rm max}(L)= c \,V \left[1 + O(V^{-1})\right]\,.
\end{equation}
On the other hand, 
$U$ is bounded as $L\to \infty$ at a continuous phase transition.
Indeed, at such transitions, in the FSS limit, any
renormalization-group invariant quantity $R$ scales as
\begin{equation}
R(\beta,L) = f_R(X) + O(L^{-\omega}) ,\quad X = (\beta-\beta_c) L^{1/\nu} ,
\label{rsca}
\end{equation}
where $f_R(X)$ is a regular function, which is universal apart from a
trivial rescaling of its argument, and $\omega$ is a 
correction-to-scaling exponent.  Therefore, $U$ has a qualitatively different
scaling behavior at a first-order or at a continuous transition. In practice,
a first-order transition can be identified by verifying that
$U_{\rm max}(L)$ increases with $L$, without the need of explicitly
observing the linear behavior in the volume.

In the case of weak first-order transitions, the nature of the
transition can also be understood from the combined analysis of $U$
and $R_{\xi}$ \cite{PV-19-CP}. At a continuous transition, in the FSS
limit the Binder parameter $U$ (more generally, any
renormalization-group invariant quantity) can be expressed in terms of
$R_\xi$ as
\begin{equation}
U(\beta,L) = F_R(R_\xi) + O(L^{-\omega}),
\label{r12sca}
\end{equation}
where $F_R(x)$ is universal. This scaling relation does not hold at
first-order transitions, because of the divergence of $U$ for
$L\to\infty$.  Therefore, the order of the transition can be
understood from plots of $U$ versus $R_\xi$.  The absence of a data
collapse is an early indication of the first-order nature of the
transition, as already advocated in Ref.~\cite{PV-19-CP}.

\section{Numerical results} \label{sec4}

\subsection{Phase behavior for the standard model with monopoles}
\label{sec4.1}

\begin{figure}[tbp]
\begin{center}
\includegraphics[width=0.6\textwidth,angle=-90]{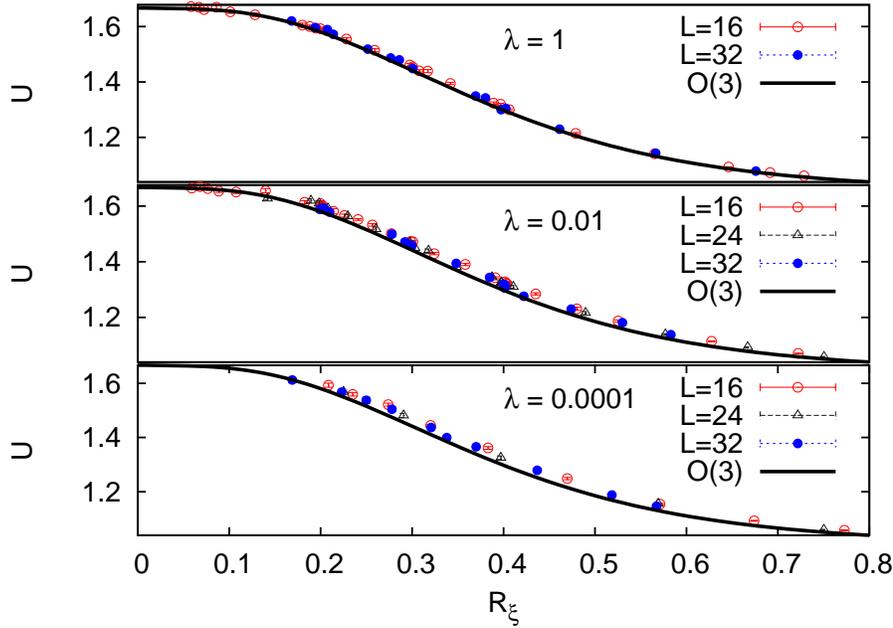}
\end{center}
\caption{Estimates of $U$ versus $R_\xi$ for the $N=2$ model (with 
  monopoles)
  for three different values of $\lambda$: $\lambda = 1$ (top panel),
$\lambda = 10^{-2}$ (middle panel), 
$\lambda = 10^{-4}$ (bottom panel). The continuous line is 
the universal curve for vector correlations 
in the O(3) vector model.
Data approach the O(3) curve as the size increases.
 }
\label{URxi-N2mono}
\end{figure}

\begin{figure}[tbp]
\begin{center}
\includegraphics*[width=0.6\textwidth,angle=-90]{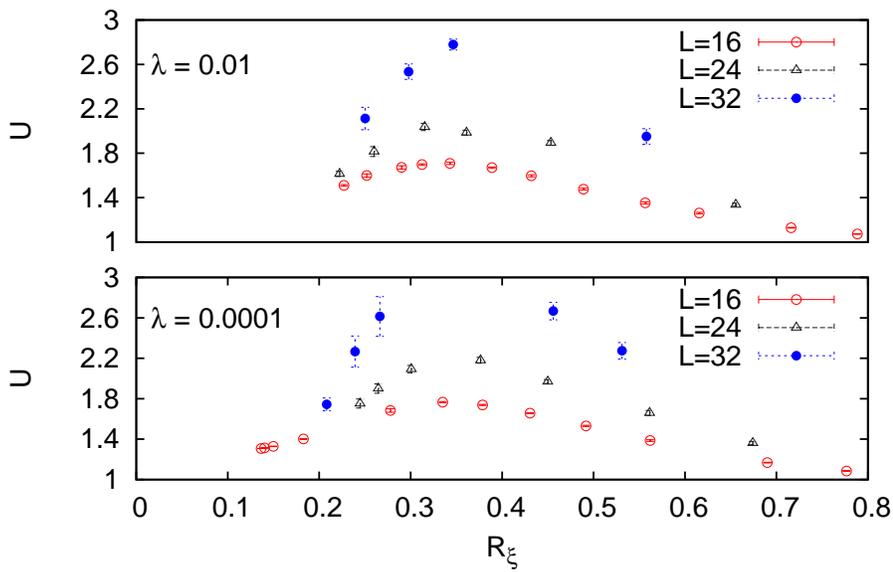}`
\end{center}
\caption{Estimates of $U$ versus $R_\xi$ for the $N=3$ model (with 
  monopoles)
  for two different values of $\lambda$: $\lambda = 10^{-2}$ (top panel), 
$\lambda = 10^{-4}$ (bottom panel). }
\label{URxi-N3mono}
\end{figure}

We have first studied the general model with Hamiltonian (\ref{Hamiltonian}),
without suppressing the monopole configurations. 
We consider systems with $N=2,3,4$, performing simulations
for several values of the parameter $\lambda$. 

For $N=2$ the transition always belongs to
the O(3) universality class, with size corrections that 
are consistent with an $1/L^\omega$, $\omega \approx 0.8$, behavior
(as expected in the O(3) case \cite{PV-02}),
and that increase at fixed $L$ as $\lambda$ decreases. 
This is supported
by the data shown in Fig.~\ref{URxi-N2mono}, where we plot the Binder 
parameter $U$ versus $R_\xi$ for three different values of $\lambda$, and 
compare the results with the universal asymptotic curve computed in the 
O(3) vector model\footnote{The data in the present model should
be compared with the scaling curve $U = F(R_\xi)$ in the
O(3) vector model, where $U$ and $R_\xi$ are computed using the
vector correlation function.
An interpolation of the O(3) data is
$F(x) = 1.666666 + x (3.0263535 + 23.139470 x)(1-e^{-15x})
- 47.838890 x^2  + 58.489668 x^3  - 67.020681 x^4  + 38.408855 x^5
- 8.8557348 x^6$. The error is smaller than 0.5\%.}. 
We observe a good agreement,
confirming that the universal critical behavior is independent of the 
potential term in the Hamiltonian. To estimate the critical point, we 
have determined the values of $\beta$ for which $R_\xi(\beta) = R^*_\xi$ and 
$U(\beta) = U^*$, where $R^*_\xi$ and $U^*$ are the universal values that 
the two quantities take at the critical point in the O(3) 
universality class. Using the estimates
\cite{Hasenbusch-20} 
$R^*_\xi = 0.564005(30)$ and $U^* = 1.13933(4)$, we obtain 
$\beta_c = 0.3082(3)$, 0.03958(2), 0.012766(1), and 0.004057(4) for 
$\lambda = 1, 10^{-2}, 10^{-3}$, and $10^{-4}$, respectively. Note that 
$\beta_c$ apparently scales as $\sqrt{\lambda}$ for $\lambda\to 0$, a behavior 
that can be understood as follows. In the mean-field approximation 
the minimum of the effective potential corresponds to $|w|^2 = 1 +
3\beta/(2\lambda)$, which indicates that $\langle |w|^2\rangle$ should 
increase as $\lambda \to 0$ roughly as $1 + 3\beta/(2\lambda) \approx
\beta/\lambda$, a behavior that is supported by the numerical data. 
If we now rescale ${\bm w} = \langle |w|^2\rangle^{1/2} {\bm z}$, the 
nearest-neighbor interaction term becomes 
\begin{equation}
    -N \hat{\beta} \sum_{x,\mu} (\bar{\bm z}_{\bm x}\cdot 
      {\bm z}_{{\bm x} + \hat{\mu}} \sigma_{{\bm x},\mu} + 
     \hbox{c.c.})  \qquad
    \hat{\beta} = \beta \langle |w|^2 \rangle
\end{equation}
We expect $\hat{\beta}_c \sim \beta_c^2/\lambda$ to be finite for 
$\lambda \to0$,  implying the expected behavior of $\beta_c$.

We have also studied the behavior for $N=3$ and 4. In both cases,
we observe a first-order transition also for finite values of $\lambda$. 
In Fig.~\ref{URxi-N3mono}, we plot the Binder parameter $U$ for 
$N=3$ and $\lambda = 10^{-2}$ and $10^{-4}$. 
It has a maximum that 
increases quite rapidly with the size of the system, 
as expected for a first-order transition. We have also analyzed the 
distributions of the order parameter $\mu_2$, observing two distinct peaks. 
Note that the identification of the transition as a first-order one becomes
easier as $\lambda$ decreases. 
The CP$^2$ $(\lambda = \infty)$
model shows a very weak first-order transition, and 
a bimodal distribution for $\mu_2$ is only observed \cite{PV-19-CP} 
for $L =  96$. 
Instead, for $\lambda = 10^{-2}$ and $10^{-4}$, the distribution
is clearly bimodal already for $L=32$.
For $N=4$ and small values of $\lambda$, 
the first-order transition is very strong
already for $L=16$: We observe large hysteresis effects and we are not 
able to obtain equilibrated results.

\subsection{Phase behavior of the monopole-free model with $N=2$}
\label{sec4.2}

\begin{figure}[tbp]
\begin{center}
\includegraphics*[width=0.6\textwidth,angle=-90]{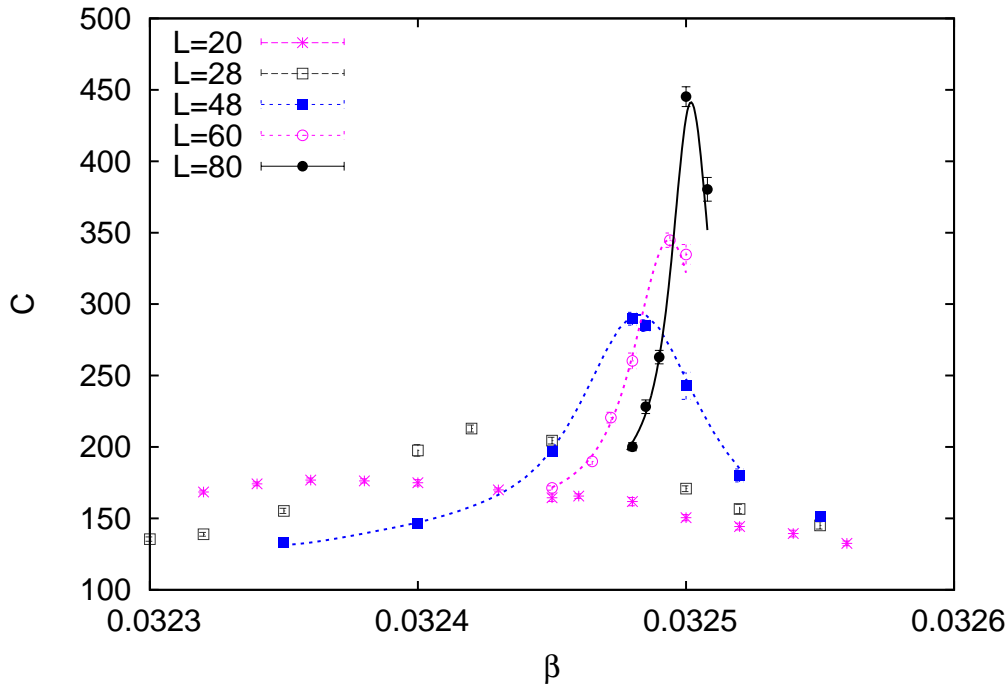}
\end{center}
\caption{Plot of the specific heat $C$ 
as a function of $\beta$ in the transition region. 
Results for several values of $L$ up to $L=80$ for the $N=2$ MF model with
$\lambda = 0.01$.
The curves interpolating the data with $L=48$, 60, and 80 are
obtained using the multihistogram reweighting method \cite{FS-89}.
}
\label{CVU-beta-N2}
\end{figure}

Let us consider the MF model for $N=2$. For $\lambda = \infty$, 
Ref.~\cite{PV-20-mfcp} was not able to 
draw any definite conclusion on the order of the
transition, in spite of  extensive simulations on lattices of
size up to $L=80$. Here we study the more general model, 
considering a very small value of $\lambda$, $\lambda = 0.01$. 
This choice is motivated by the results of Section~\ref{sec4.1}. 
If the transition for $\lambda = \infty$ is continuous, we expect 
the simulations for $\lambda = \infty$ and $\lambda = 0.01$ to provide 
consistent results. 
This universality check would support the existence of a MF universality 
class for $N=2$. On the other hand, if no such universality class exists 
and the transition is of first order, 
one might hope the first-order nature of the 
transition to become more evident as $\lambda$ decreases, as it occurs in the 
presence of monopoles. 

In Fig.~\ref{CVU-beta-N2} we report the specific heat $C$ 
as a function of $\beta$. It has a maximum for $\beta \approx 0.0325$
which increases with the size of the lattice.
For each value of $L$ we have determined $C_{\rm max}(L)$. 
We have fitted the results to $a L^\delta$ including only data with 
$L\ge L_{\rm min}$, obtaining $\delta = 0.58(1)$, 0.64(1), 0.68(1), 0.81(5) for 
$L_{\rm min} = 16,20,28,48$. If we perform a fit to 
$a L^\delta + b$, i.e., if we include an analytic correction, we obtain 
$\delta = 0.86(4), 0.88(6), 1.01(10)$ for
$L_{\rm min} = 16,20,28$.
The results are analogous to those
obtained for $\lambda=\infty$. Ref.~\cite{PV-20-mfcp}
obtained $\delta = 0.8(1)$ from the analysis of the specific heat 
using only data satisfying  $L\ge 48$.
The exponent $\delta$ is very 
different from the one expected for a first-order
transition, $\delta = 3$. 
If the transition is continuous, assuming the hyperscaling relation
$2 - \alpha = 3 \nu$, 
we can estimate $\nu$ using $\nu = 2/(3 + \delta)$. If $\delta = 1.0(2)$,
we would obtain $\nu = 0.50(2)$, which is compatible 
with the estimate \cite{PV-20-mfcp} $\nu = 0.52(2)$ for the MFCP$^1$ model.

\begin{figure}[tbp]
\begin{center}
\includegraphics*[width=0.6\textwidth,angle=-90]{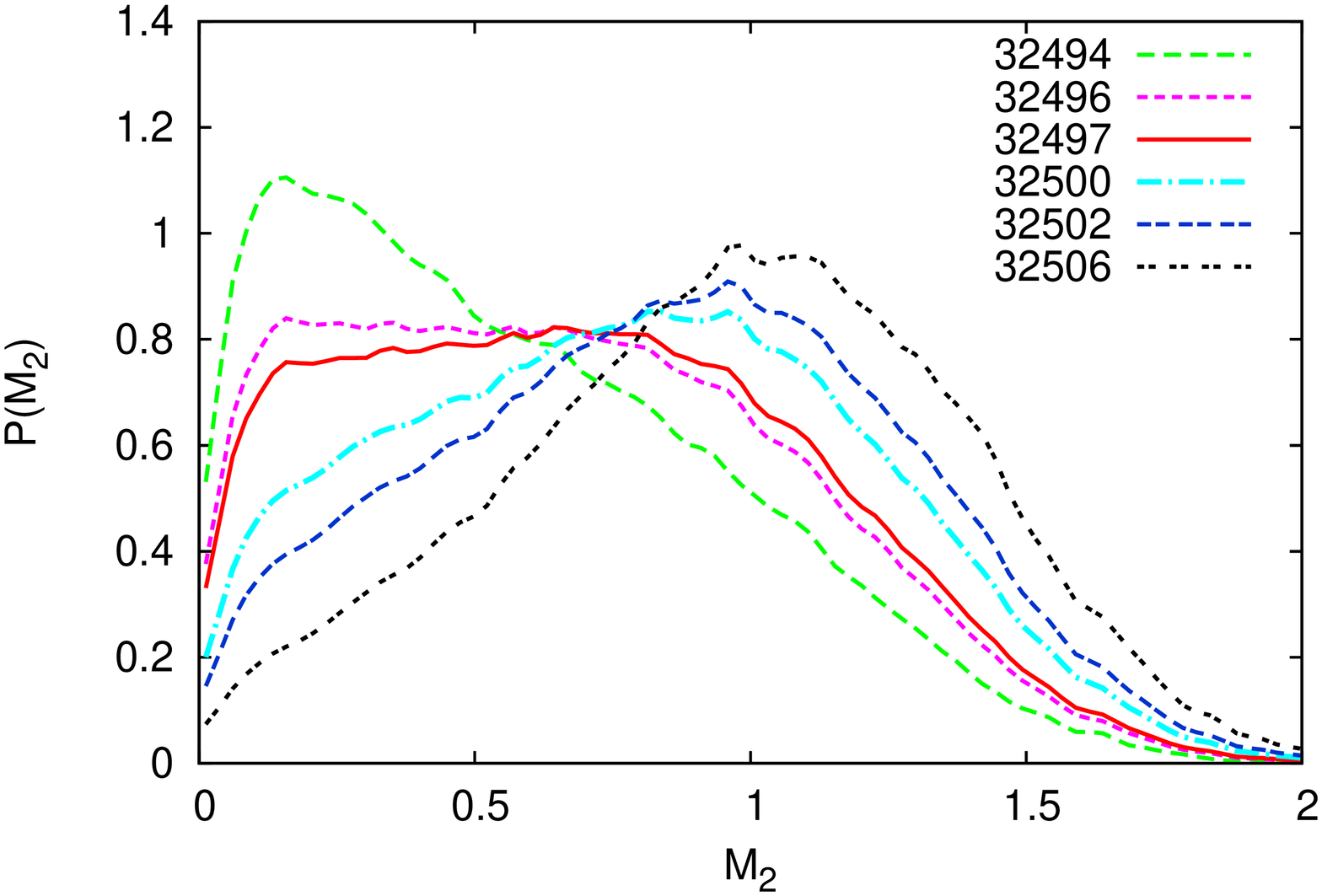}
\includegraphics*[width=0.6\textwidth,angle=-90]{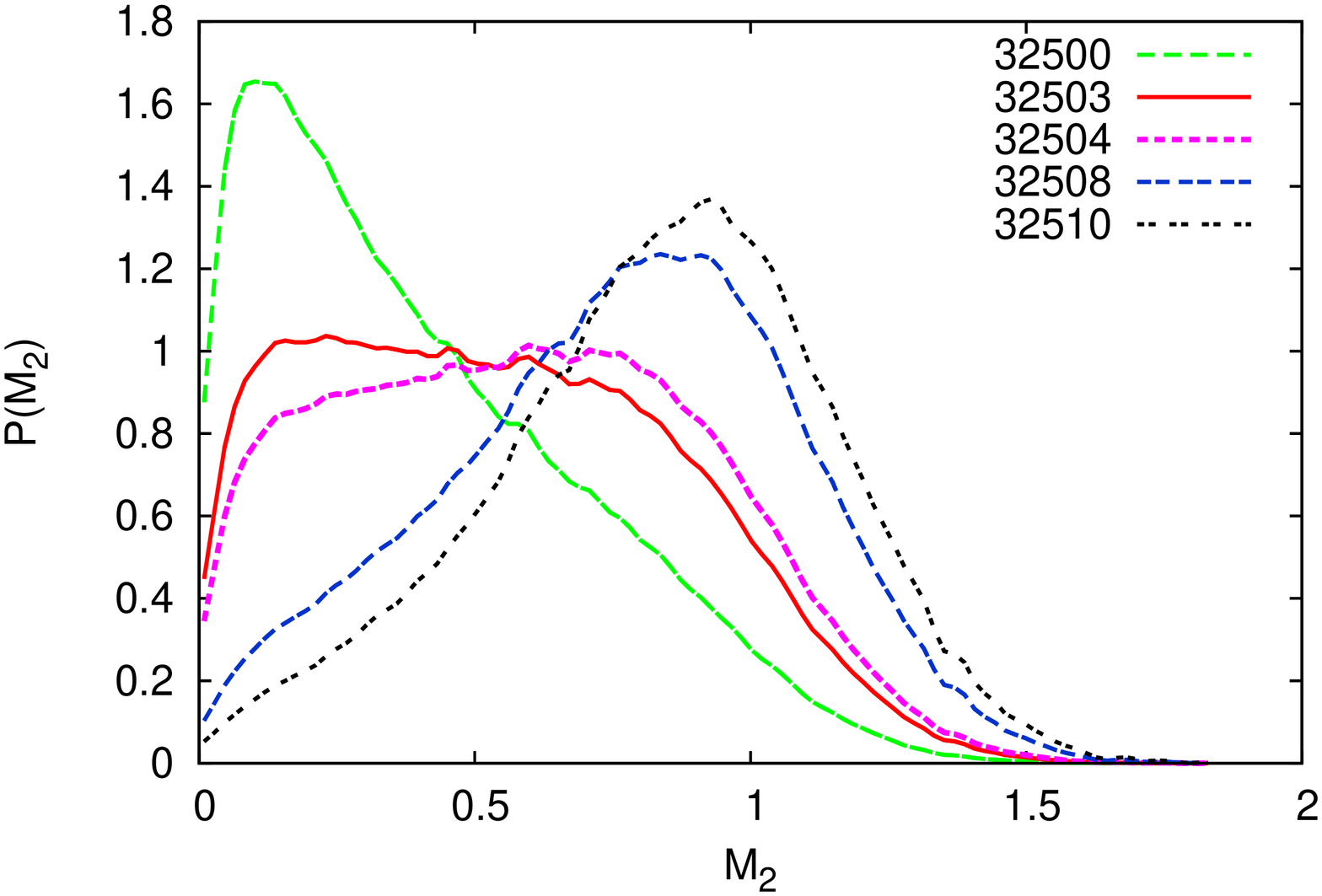}
\end{center}
\caption{Distribution $P(M_2)$ for the $N=2$ MF model with $\lambda = 0.01$
  for $L=60$ (top) and $L=80$ (bottom) for 
  different values of $\beta$.  The distributions are obtained
  using the multihistogram reweighting method \cite{FS-89}. 
  In the legend we report $10^6 \beta$; e.g., 32494 corresponds to 
  $\beta = 0.032494$. }
\label{PM2-N2}
\end{figure}

\begin{figure}[tbp]
\begin{center}
\includegraphics*[width=0.6\textwidth,angle=-90]{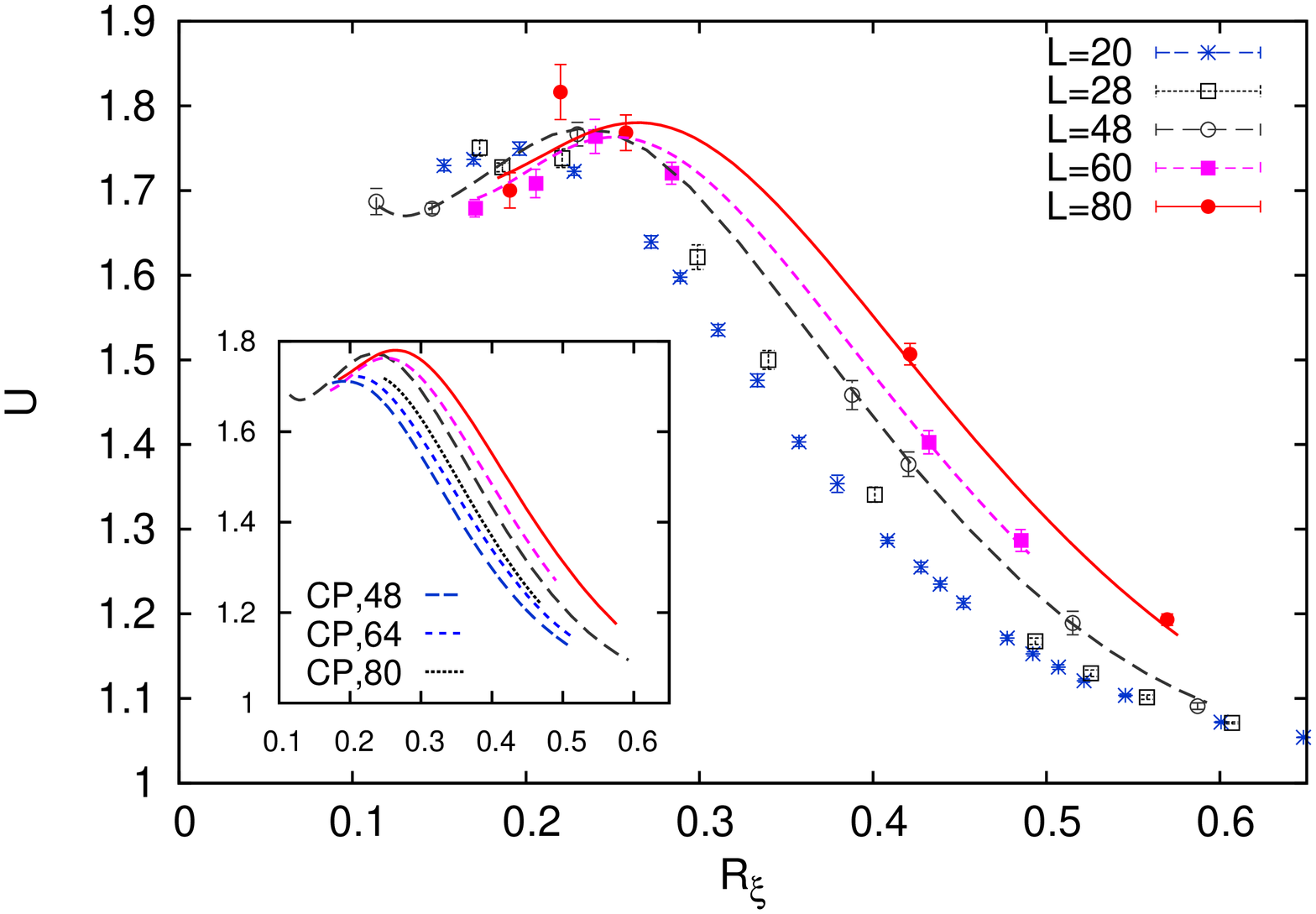}
\end{center}
\caption{Plot of the Binder parameter $U$ versus $R_\xi$, for several
  values of $L$ up to $L=80$ for the $N=2$ MF model with $\lambda = 0.01$. 
  The curves interpolating the
  data with $L=48$, 64, and 80 ($L$ increases moving rightward) are
  obtained using the multihistogram reweighting method \cite{FS-89}.
  In the inset we report the same interpolating curves (three rightmost 
  curves) together with the analogous interpolating curves 
  (three leftmost curves, again $L$ increases moving rightward) 
  obtained for $\lambda=\infty$ (MFCP$^1$ model). }
\label{URxi-N2}
\end{figure}

To better understand the nature of the transition, we have determined the 
distribution of $\mu_2$ defined in Eq.~(\ref{binderdef}),
\begin{equation}
P(M_2) = \langle \delta(M_2 - \mu_2) \rangle.
\end{equation}
It is reported in Fig.~\ref{PM2-N2} for several values of $\beta$. 
Although no double-peak structure
is observed, the distribution varies 
as expected for a first-order transition, with 
a sharp change of the position of the peak as $\beta$ is varied. 
Finally, in Fig.~\ref{URxi-N2} we report $U$ versus $R_\xi$. 
As already observed for $\lambda=\infty$, data do not scale. At
fixed $R_\xi$, the estimates of $U$ are systematically increasing with
$L$ for $0.2\le R_\xi\lesssim 0.6$. However, we do not observe any
systematic increase of the maximum of $U$ with $L$, as expected for 
a first-order transition. Precisely,
using the multihistogram reweighting method \cite{FS-89}, we estimate
$U_{\rm max}(L) = 1.77(1)$, 1.76(2), and 1.78(2) for $L=48,60$, and 80. 
It is interesting to compare the behavior of $U$ for $\lambda = \infty$ 
(MFCP$^1$ model; results from Ref.~\cite{PV-20-mfcp}) 
and $\lambda = 0.01$ (see the inset of Fig.~\ref{URxi-N2}). The curves 
for the two different values of $\lambda$ are apparently unrelated. 
There is really no indication for universality, making the 
continuous-transition scenario rather unlikely.


\begin{figure}[tbp]
\begin{center}
\includegraphics*[width=0.6\textwidth,angle=-90]{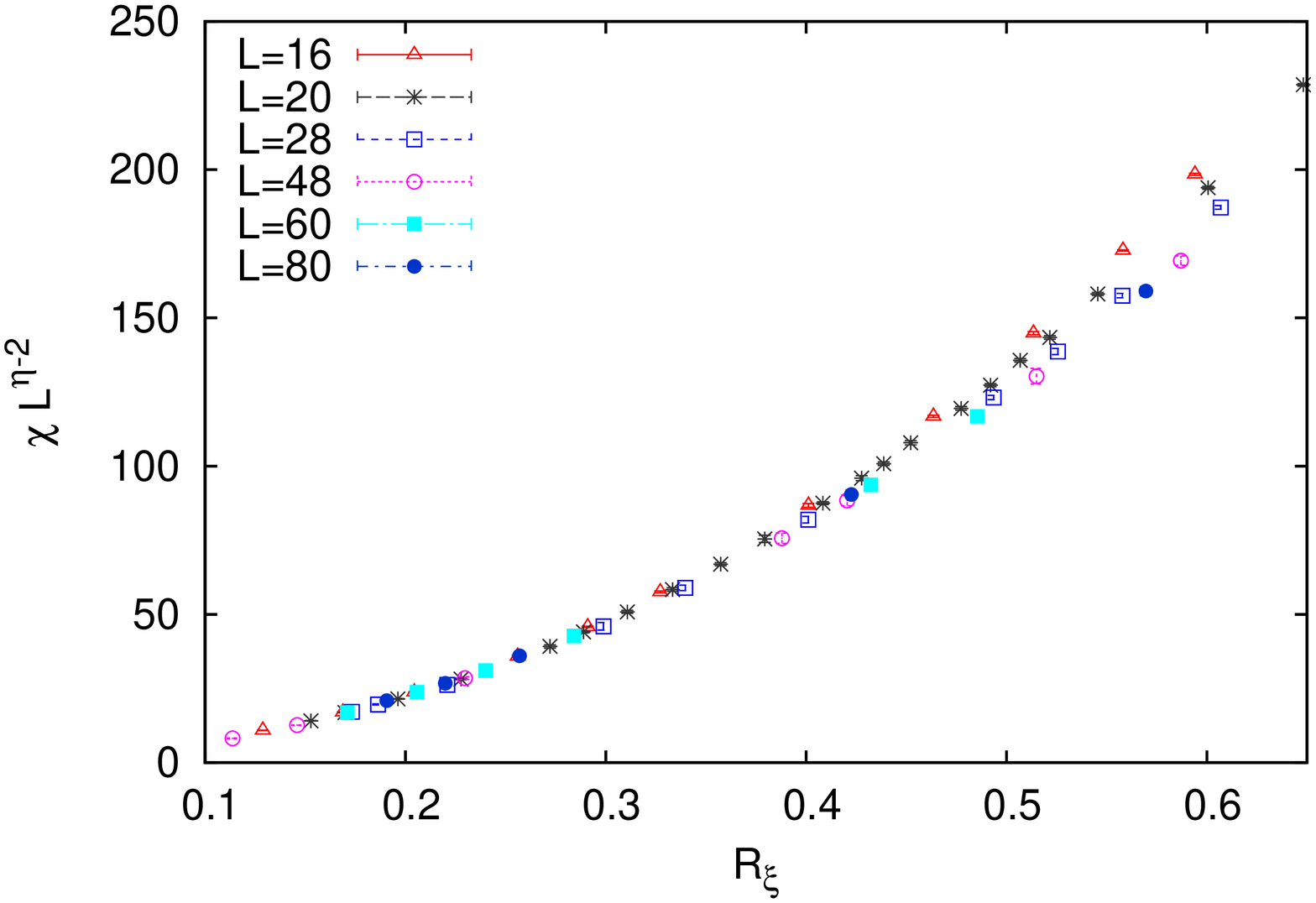}
\end{center}
\caption{Plot of $\chi L^{\eta-2}$ as a function of $R_{\xi}$ 
with $\eta = 0.223$. Results for the $N=2$ MF model with $\lambda = 0.01$.}
\label{scalingplotchi-N2nomono}
\end{figure}

To conclude the analysis of the available data, we may assume that the
transition is continuous and determine the critical exponents. We have first
performed fits of $R_\xi$ and $U$ without including scaling corrections.
The exponent $\nu$ and the transition value $\beta_c$ 
are determined by fitting the data
to Eq.~(\ref{rsca}). The function $f_R(x)$ is approximated by a
polynomial. The fits of $R_\xi$ have a large $\chi^2$, unless only 
data with $L\ge 48$ are included. In this case we obtain 
$\beta_c = 0.0325086(9)$ and $\nu= 0.54(2)$. If we use the same set of data 
for the Binder parameter, we obtain $0.0325114(7)$ and $\nu= 0.43(1)$.
The estimates of $\beta_c$ and of the exponent obtained from the 
two quantities are clearly not consistent. To understand whether 
these discrepancies are due to scaling corrections, we have performed 
combined fits of $R_\xi$ and $U$ to 
\begin{equation}
R(\beta,L) = f_R(X)  + L^{-\omega} g_R(X),
\end{equation}
parametrizing $f_R(X)$ and $g_R(X)$ with polynomials. As before 
$X = (\beta - \beta_c)L^{1/\nu}$. Fitting all data with $L\ge 16$, we obtain
$\omega = 0.1(1)$, $\nu = 0.50(1)$, $\beta_c = 0.32513(1)$. If we only
include data with $L\ge 20$, we obtain consistent estimates for all quantities
and $\chi^2/\hbox{DOF} \approx 1$ (DOF is the number of degrees of freedom of
the fit). While the estimates of the exponents and of $\beta_c$ are reasonable,
the estimates of the asymptotic curves $f_R(X)$ have large errors
and, in some cases, they violate rigorous inequalities. For instance, 
the scaling curve
$f_U(X)$ for the Binder parameter $U$ 
should always satisfy $f_U(X) > 1$. This bound is 
violated for $X \gtrsim 0$. For instance, the fit predicts
$f_U(0) = 0.6(2)$. Therefore, the results of these fits cannot be trusted. 
This is not unexpected. If $\omega$ is so small, next-to-leading scaling
corrections with exponents $2\omega$, $3\omega$, $\ldots$, should 
be included as well, to obtain meaningful results.

For $\lambda = \infty$ the best evidence \cite{PV-20-mfcp} 
for a continuous transition was
provided by the scaling behavior of the susceptibility $\chi$ defined in
Eq.~(\ref{chisusc}). This quantity scales as
\begin{equation}
\chi(\beta,L) \sim L^{2-\eta} \left[ f_\chi(X) + O(L^{-\omega})\right],
\label{chisca}
\end{equation}
or, equivalently, as
\begin{equation}
\chi(\beta,L) \sim L^{2-\eta} \left[ F_\chi(R_\xi) + O(L^{-\omega})\right].
\label{chisca2}
\end{equation}
For $\lambda = \infty$, data follow the scaling behavior (\ref{chisca2}) 
quite precisely
with $\eta = 0.335(10)$. We have repeated the same analysis using 
the present results
for $\lambda = 0.01$.
We fit $\chi$ to $\ln \chi = (2 - \eta) \log L +
\hat{f}_\chi(R_\xi)$, where we approximate the function
$\hat{f}_\chi(x)$ with a polynomial in $x$.  To estimate the role of
the scaling corrections we include in the fit only the data
corresponding to sizes $L\ge L_{\rm min}$. We obtain $\eta = 0.239(6)$
and 0.208(17) for $L_{\rm min} = 20$ and 28, respectively. In this
case, scaling corrections appear to be small ($\chi^2$/DOF is approximately
1.4 for $L_{\rm min} = 20$ and 0.1 for $L_{\rm min} = 28$, if 
only data satisfying $R_\xi \le 0.45$ are considered; DOF 
is the number of degrees of freedom of the fit), as is also evident from
the scaling plot, Fig.~\ref{scalingplotchi-N2nomono} (in the figure 
we use $\eta = 0.223$, which is the average of the results obtained for 
the two values of $L_{\rm min}$). Although results are 
consistent with the behavior expected at a continuous transition, note 
that universality is strongly violated: The estimate of $\eta$ differs
from the one obtained in the MFCP$^1$ model
\cite{PV-20-mfcp}, $\eta = 0.335(10)$. It is closer to, though not in agreement with, the 
estimate \cite{NCSOS-15} $\eta = 0.259(6)$, obtained in a loop
model that is expected \cite{NCSOS-11,NCSOS-13} to belong to the same 
universality class.

In conclusion, our results are apparently not consistent with a scenario 
in which the MF model with $N=2$ undergoes a continuous transition,
unless scaling corrections decay with a very small exponent or logarithmically,
as suggested in Ref.~\cite{Sandvik-10}. In our view, 
the most likely possibility is that the transition 
is of first order, but so weak that a clear signature can only be obtained 
on significantly larger lattices.

\subsection{Results for $N=25$} \label{sec4.3}

\begin{figure}[tbp]
\begin{center}
\includegraphics*[width=0.6\textwidth,angle=-90]{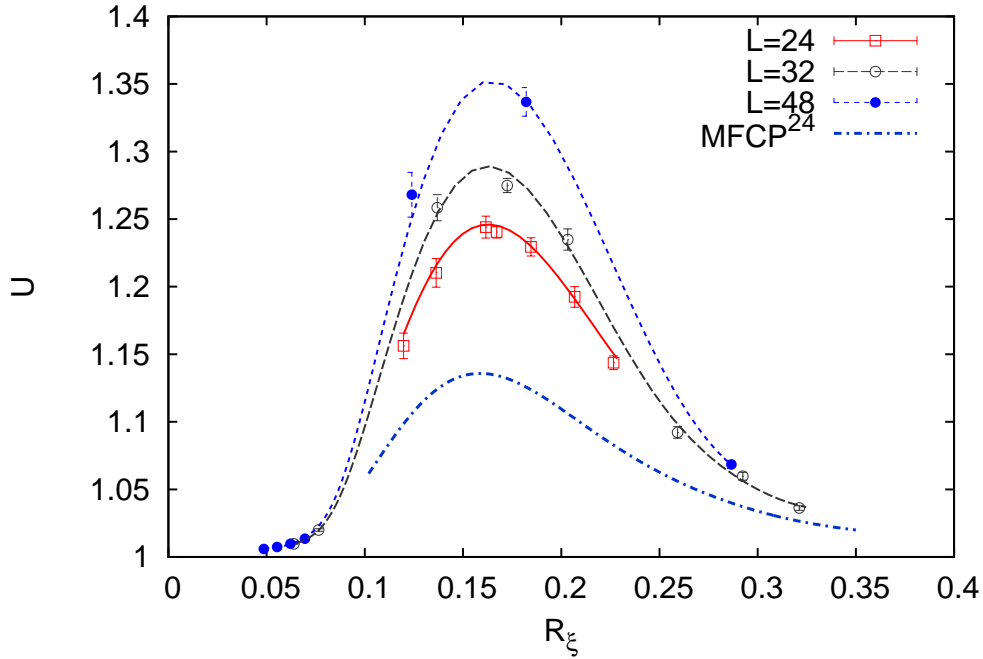}
\end{center}
\caption{Plot of the Binder parameter $U$ versus $R_\xi$, for several
  values of $L$ for the MF model with $N=25$ and $\lambda = 100$.  The curves
  interpolating the
  data are obtained
  using the multihistogram reweighting method \cite{FS-89}. 
  We also report an extrapolation (lower dot-dash line) 
  of the data for the $\lambda=\infty$ model 
  (MFCP$^{24}$ model). }
\label{UvsRxi-N25-l100}
\end{figure}

\begin{figure}[tbp]
\begin{center}
\includegraphics*[width=0.6\textwidth,angle=-90]{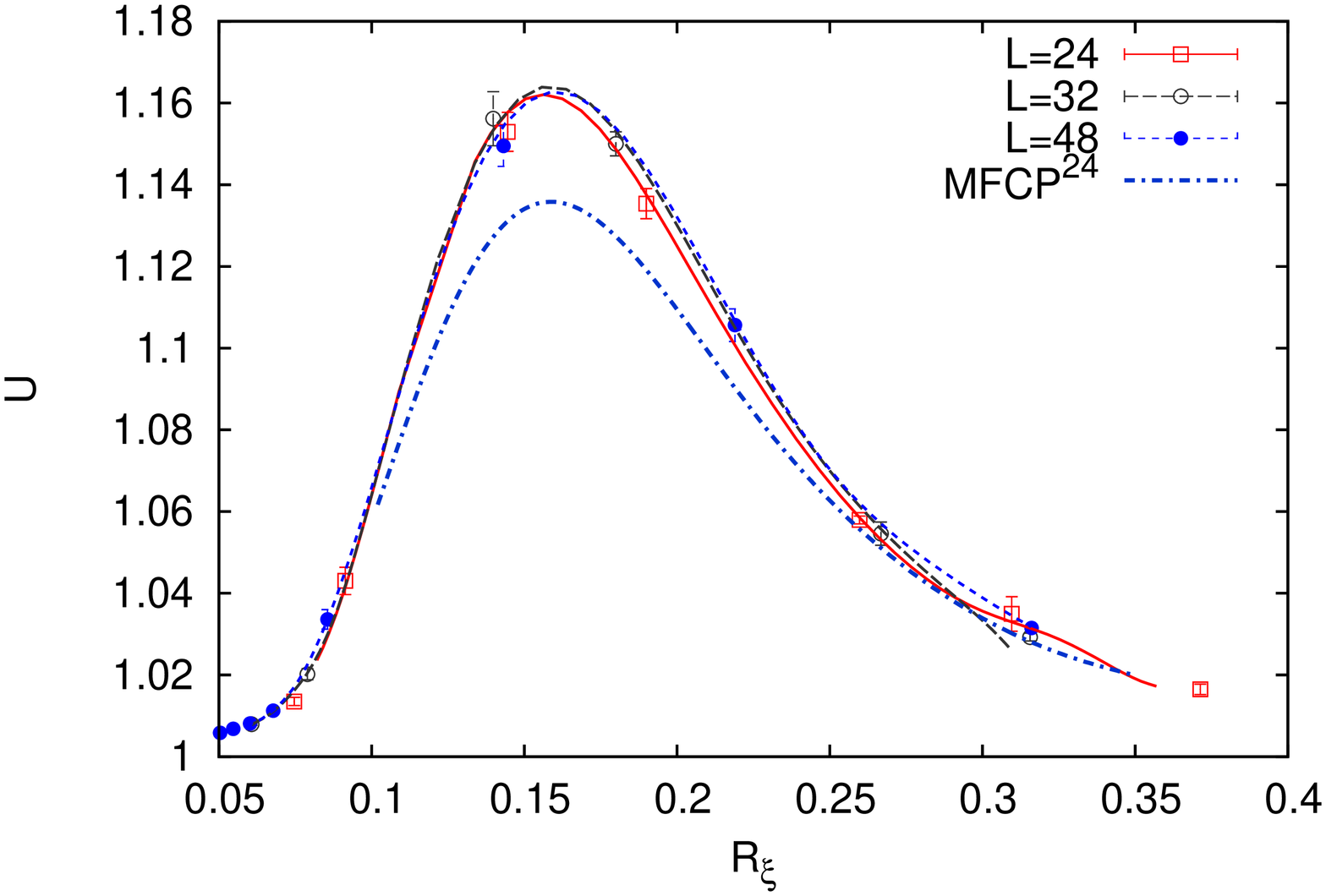}
\end{center}
\caption{Plot of the Binder parameter $U$ versus $R_\xi$, for several
  values of $L$ for $N=25$ for the model with partition function 
  (\ref{Z-bistabile}) and $a = \alpha = 0.96$.  The curves
  interpolating the data are obtained
  using the multihistogram reweighting method \cite{FS-89}. 
  We also report an extrapolation (lower dot-dash line) 
  of the data for the $\lambda=\infty$ model (MFCP$^{24}$ model). }
\label{UvsRxi-N25-bist}
\end{figure}

We finally present our results for $N=25$. We have performed
simulations on lattices of size $16\le L \le 48$ for several values of 
$\lambda$. For $\lambda = 0.01$, 1, and 10, we observe a very strong
first-order transition. If we first increase $\beta$ and then decrease it
across the transition, a strong hysteresis is observed, even for $L$ 
as small as 12. For $\lambda = 100$, data are again consistent with 
a first-order transition. In Fig.~\ref{UvsRxi-N25-l100}, we plot 
$U$ versus $R_\xi$. Data do not scale and the maximum $U_{\rm max}(L)$
increases with $L$, which is the typical signature of a first-order 
transition. In the same figure we also report an extrapolation
of the data \cite{PV-20-mfcp} obtained for the MFCP$^{24}$ model 
which is obtained in the limit $\lambda=\infty$.
Also in the MFCP$^{24}$ model 
$U$ has a maximum, which is however significantly smaller
than the maxima observed for $\lambda = 100$. Clearly, the deviations observed
for $\lambda = 100$ cannot be interpreted as corrections to scaling. 

To further explore the role that the size fluctuations of the field play, we 
have studied a different model that is easier to simulate and in which 
field-size fluctuations can be easily controlled. In the modified model the 
partition function is given by 
\begin{equation}
Z = \int \prod_{{\bm x}\mu} d \sigma_{{\bm x}\mu} 
        \prod_{{\bm x}} [d{\bm w}_{{\bm x}} d\bar{\bm w}_{{\bm x}}  
        M({\bm w}_{\bm x})] \,
        e^{-\beta H_{\rm kin}}
\label{Z-bistabile}
\end{equation}
where $H_{\rm kin}$ is defined in Eq.~(\ref{hcpnla}) and 
\begin{equation}
    M({\bm w}) = |w|^{1-2N} [\delta(|{w}| - 1) + 
                             \alpha \delta(|{w}| - a)].
\end{equation}
In this model $|{w}|$ takes only two values, 1 and $a$. 
Field-size fluctations can 
be controlled by changing the parameters $a$ and $\alpha$. At $\beta = 0$,
the probabilities $P(|w|)$ of the two values are given by
$P(1) = \cal N$ and $P(a) = \alpha {\cal N}$ with ${\cal N}=1+\alpha$. 
If $\beta$ is turned on, 
the probability $P(a)$ decreases with increasing $\beta$, since the 
kinetic term favors configurations with $|w| = 1$. 
We performed simulations for $a = \alpha = 0.96$, observing a transition for 
$\beta\approx 0.324$.  In the critical region, we find 
$\langle w^2 \rangle \approx 0.985$, so that $P(1) \approx 0.80$ and 
$P(a) \approx 0.20$.
To compare the results obtained for this model with those 
obtained at fixed $\lambda$, we can compare the width of the fluctuations,
defining
\begin{equation}
\sigma^2_{w^2} = { \langle (|{w}|^2 - \langle |{w}|^2\rangle)^2 
                   \rangle \over 
                   \langle |{w}|^2\rangle^2}.
\end{equation}
In the model at fixed $\lambda = 100$, we obtain $\sigma_{w^2} \approx 0.057$ 
in the whole critical region for $L=48$. In the new model, for 
the same value of $L$ we obtain $\sigma_{w^2} \approx 0.031$. 
Thus, in these simulations the fluctuations of $|{\bm w}|^2$ are reduced by
a factor of 2. 

\begin{figure}[tbp]
\begin{center}
\includegraphics*[width=0.6\textwidth,angle=-90]{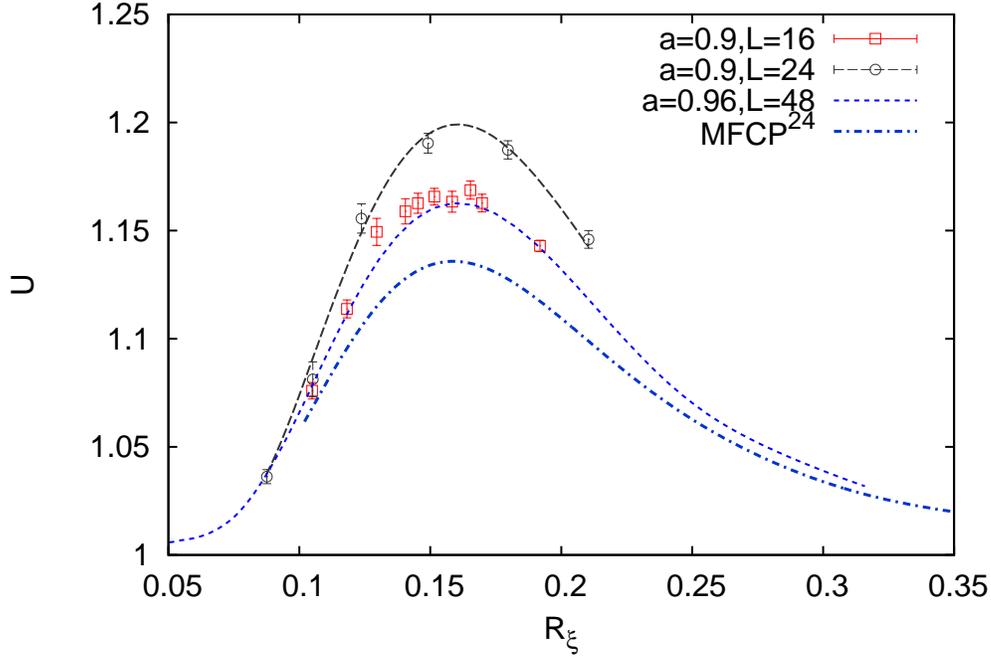}
\end{center}
\caption{Plot of the Binder parameter $U$ versus $R_\xi$, for several
  values of $L$ for $N=25$ for the model with partition function 
  (\ref{Z-bistabile}). We report: data obtained for the model
  with  $a = 0.9$ and $\alpha = 2$ ($L=16, 24$) and the interpolation 
  (long dash; upper curve) of the 
  $L=24$ data using the multihistogram reweighting method \cite{FS-89}; 
  the interpolation (short dash; middle curve) of the data with $L=48$ 
for the model with 
  $a = \alpha = 0.96$; the  extrapolation (dot dash; lower curve) 
   of the data for the $\lambda=\infty$ model (MFCP$^{24}$ model). }
\label{UvsRxi-N25-bist2}
\end{figure}

The results for $a = 0.96$ are consistent with a continuous transition, since
we observe good scaling when we plot $U$ versus $R_\xi$, see 
Fig.~\ref{UvsRxi-N25-bist}. The results for different values of $L$, obtained
using the multihystogram method \cite{FS-89}, essentially fall on top of each
other. 
However, note that the data scale onto a curve 
that is different from the one observed in 
the model at $\lambda = \infty$. For instance, in the present model 
$U_{\rm max} (L) = 1.164(5)$, 1.163(5) for $L=32,48$, respectively.
On the other hand, for the fixed-length model with $\lambda = \infty$, 
$U_{\rm max} (L) \approx 1.136$, see Fig.~\ref{UvsRxi-N25-bist}. Given the stability of the results, 
such large difference does not appear to be the result of scaling corrections,
unless the subleading exponent is very small.

We have determined the 
exponent $\nu$ as before, fitting $U$ and $R_\xi$ to
Eq.~(\ref{rsca}) using a polynomial approximation for the scaling function
$f_R(x)$.  We obtain:
\begin{eqnarray} 
\beta_c = 0.32471(2), \qquad
\nu = 0.59(2), \quad && \hbox{from $U$;} \nonumber \\
&& \\
\beta_c = 0.324675(15), \qquad
\nu = 0.57(1), \quad && \hbox{from $R_\xi$.}  \nonumber 
\end{eqnarray}
The estimates of the exponent and of $\beta_c$ obtained 
from the analysis of these 
two quantities are in substantial agreement. Moreover, the estimates of $\nu$
are consistent with the estimate $\nu = 0.595(15)$ obtained in
Ref.~\cite{PV-20-mfcp}. 

Finally, we study the critical behavior of the susceptibility
$\chi$, performing fits to the ansatz
\begin{equation}
\ln \chi = (2 - \eta) \ln L +
\hat{f}_\chi(R_\xi). 
\end{equation}
 We obtain $\eta = 0.945(15), 0.93(2)$
for $L_{\rm min} = 24,32$, respectively, 
with a very good collapse of the data.  
However, these estimates are not consistent with
the result $\eta = 0.87(1)$ obtained in Ref.~\cite{PV-20-mfcp}. Again,
universality is not observed.

To conclude, we have studied the same model, setting $a = 0.9$ and 
$\alpha = 2$, to increase the size of the field-length fluctuations.
In the critical region, for $L=24$, we obtain 
$\langle |w|^2 \rangle \approx 0.978$,
$\sigma_{w^2} \approx 0.047$, 
$P(1) \approx 0.88$, and $P(a) \approx 0.12 $. 
The plot of $U$ versus $R_\xi$ is shown in Fig.~\ref{UvsRxi-N25-bist2}, for two
values of $L$. In this case, data do not scale. The maximum 
$U_{\rm max}(L)$ of the Binder parameter increases with $L$, which provides 
some indication that the transition is not continuous.

In conclusion, the results we have obtained are not consistent with 
universality and are better explained in terms of a discontinuous 
transition, that becomes weaker as field-length fluctuations decrease. 
If we trust the results of Ref.~\cite{PV-20-mfcp}, i.e., if 
we assume that the transition in the MFCP$^{24}$  model is continuous, 
we conclude that field-length fluctuations 
are relevant perturbations of the 
$N=25$ MF universality class. This implies that this critical behavior
cannot have a simple field-theory
interpretation. Indeed, in the field-theory approach the strength of the
quartic potential parameters is always irrelevant. Of course, it is 
equally possible that no MF universality class exists. In that case 
the transition in  the MFCP$^{24}$  model would be discontinuous. However,
the correlation length at the transition would be finite but so large 
($\xi\gtrsim 80$) 
that an apparent scaling behavior would be observed in the simulations
with $L\le 80$
as if the transition were continuous. 

\section{Conclusions} \label{sec5}

This paper reports a study of the phase diagram  and of 
the nature of the phase transitions of 3D lattice models characterized by
a global SU($N$) symmetry and a local U(1) symmetry.  They generalize
the lattice nearest-neighbor 
CP$^{N-1}$ model with an explicit gauge field---the
corresponding Hamiltonian is given in Eq.~(\ref{hcpnla}). 
At variance with the CP$^{N-1}$ model,
the length of the fields is not fixed,
but is controlled by adding an appropriate potential parametrized
by a parameter $\lambda$: for $\lambda\to \infty$ we reobtain the 
CP$^{N-1}$ model.
We study the role that monopoles play in this class of systems, restricting
the configuration space to gauge-field configurations in which no
monopoles (we use the definition of Ref.~\cite{DGT-80}) are present. 
We perform Monte Carlo
simulations for $N=2$ and 25, with the purpose of comparing the 
results for the present model with those obtained in
Ref.~\cite{PV-20-mfcp} for the MFCP$^{N-1}$ model.  
The analysis of the 
finite-size data allows us to identify a 
finite-temperature transition in all cases, related to the
condensation of a local gauge-invariant bilinear order parameter $Q_{\bm x}$,
defined in Eq.~(\ref{qdef}).

For $N=2$ we performed simulations on relatively large systems, 
up to $L=80$. As it occurred for $\lambda = \infty$, 
we are not able to draw a definite
conclusion on the nature of the transition. 
Many features are typical of first-order transitions. For instance,
the Binder parameter data do not scale when
plotted versus $R_\xi = \xi/L$, and the distributions of the order
parameter and of the energy are quite broad, although without the
typical bimodal shape that signals the presence of coexisting
phases. On the other hand, the maximum of the Binder parameter does not 
increase  with $L$, as it would be expected at first-order transitions. 
Assuming the 
transition to be continuous, we also estimate some critical  exponents.
The exponent $\nu$ is estimated from the finite-size behavior of 
$R_\xi$ and $U$. The fits, however, provide inconsistent estimates. Fits of the 
susceptibility give $\eta = 0.22(2)$. The quality of the fit is 
good and we observe a good collapse of the data when $\chi L^{2-\eta}$ is 
plotted versus $R_\xi$. However, the estimate of $\eta$ differs significantly
from the MFCP$^1$ result \cite{PV-20-mfcp} $\eta = 0.335(10)$, 
in contrast with universality.

Taking all results into account, the simplest scenario that explains 
the observed behavior is that of a weak first-order transition. 
It is so weak that coexisting phases can only be observed on very 
large lattices with $L\gg 80$, because of the presence of a large 
effective scale induced by the no-monopole condition. Of course, 
we cannot exclude,  as suggested in the literature on quantum
antiferromagnets, that the transition is continuous with 
very slowly decaying---even logarithmic \cite{Sandvik-10},
associated with a
dangerously irrelevant variable---scaling corrections.

Finally, we have studied the MF model with $N=25$. We have performed
simulations for several values of the potential parameter $\lambda$. 
For $\lambda = 0.01,1$ and 10 we observe a strong first-order transition
and a bimodal order-parameter distribution 
is already observed for lattice sizes as small as 
$L=12$. For $\lambda = 100$, we do not observe a two-peak structure in the 
distributions of the energy or of the order parameter $\mu_2$ defined in 
Eq.~(\ref{binderdef}). However, the rapid increase of the maximum 
of the Binder parameter with $L$ clearly points towards a 
discontinuous transition.  We have also studied a different model
in which the field length can only take two different values. 
In both cases, we find that the results for the two generalized models are not
consistent with those obtained in the MFCP$^{24}$ model. They are better 
explained by a first-order transition scenario. These results cast doubts on 
the existence of a MF universality class for $N=25$. In any case, note that,
if such universality class really exists, it cannot have a simple 
field-theory interpretation, since field-length fluctuations are apparently 
relevant perturbations in the renormalization-group sense. We do not think 
additional numerical work can provide new insight on these issues. A more
promising approach is probably the $1/N$ expansion, in which some answers
can be obtained analytically for large values of $N$.

\end{document}